\documentclass{article}
\usepackage{graphicx} % Required for inserting images
\usepackage[english]{babel}
\usepackage[letterpaper,top=2cm,bottom=2cm,left=3cm,right=3cm,marginparwidth=1.75cm]{geometry}
\usepackage{authblk}
\usepackage{blindtext}
\usepackage{cite}

\usepackage{amsmath}
\usepackage{amssymb}
\usepackage{graphicx}
\usepackage[colorlinks=true, allcolors=blue]{hyperref}
\usepackage{tikz-feynman}

\newcommand{\dd}{\text{d}}

\newcommand{\hs}{\hspace{0.5cm}}
\newcommand{\pr}{^{\,\prime}}
\newcommand{\J}{\mathcal{J}}
\renewcommand{\O}{\mathcal{O}}

\usepackage{environ}
\usepackage{tikz}
\tikzset{Witten diagram/.style={execute at begin picture={%
\draw[blue,fill=white!20] circle[radius=\pgfkeysvalueof{/tikz/Witten/radius}];
}},vertex/.style={circle,fill,inner sep=1.5pt,node
contents={}},
Witten/.cd,radius/.initial=2cm}
\NewEnviron{wittendiagram}[1][]{\vcenter{\hbox{\begin{tikzpicture}[Witten diagram,#1]%
\BODY
\end{tikzpicture}}}}

\tikzset{sWitten diagram/.style={execute at begin picture={%
\draw[blue,fill=white!20] circle[radius=\pgfkeysvalueof{/tikz/sWitten/radius}];
}},vertex/.style={circle,fill,inner sep=1.0pt,node
contents={}},
sWitten/.cd,radius/.initial=1cm}
\NewEnviron{swittendiagram}[1][]{\vcenter{\hbox{\begin{tikzpicture}[sWitten diagram,#1]%
\BODY
\end{tikzpicture}}}}

\title{AdS amplitudes as CFT correlators}

\author[1]{Máximo Bañados\thanks{maxbanados@gmail.com}}
\author[2]{Ernesto Bianchi\thanks{E.Bianchi@soton.ac.uk}}
\author[1]{Iván Muñoz\thanks{ivanmunozsandoval@gmail.com}}
\author[2]{Kostas Skenderis\thanks{K.Skenderis@soton.ac.uk}}
\affil[1]{Facultad de Física, Pontificia Universidad Católica de Chile, Santiago, Chile}
\affil[2]{STAG Research Centre \& Mathematical Sciences,  University of Southampton, Highfield, SO17 1BJ Southampton, UK}

\begin{document}

\maketitle

\begin{abstract}
We show that AdS amplitudes are CFT correlators to all orders in the loop expansion by showing that they obey the conformal Ward identities. In particular, we provide explicit formulas for the constants and functions of cross-ratios that determine the CFT correlators in terms of bulk data.

\end{abstract}

%\tableofcontents

\section{Introduction}

In a theory of quantum gravity there are no bulk local invariants (because diffeomorpshims act on spacetime points). In spacetimes with asymptotia, we need to impose boundary conditions at infinity, and one may define local operators at the (conformal) boundary via the boundary conditions. For example, one may require that a bulk scalar field takes a prescribed value at the boundary. The gravitational path-integral 
computed with such boundary conditions would then compute observables that depends on boundary points. Such observables may be organised according to their transformation properties under the asymptotic symmetry group, the group of transformations that preserves the boundary conditions. 

In the case of asymptotically (locally) AdS gravity, the boundary carries a conformal structure, so this construction naturally produces $n$-point functions, which we will call AdS amplitudes, that transform as CFT correlators. The AdS/CFT conjectures \cite{JM1, Gubser:1998bc, Witten} asserts that AdS gravity is equivalent to a local CFT in one dimension less and in particular AdS amplitudes are equal to CFT correlators\footnote{This perspective on the duality has been emphasised early on in \cite{Giddings:1999qu}.}. The relation between AdS amplitudes computed via Witten diagrams and CFT correlators has been tested with tree-level examples already in the foundational papers \cite{Gubser:1998bc, Witten} and numerous explicit evaluations of AdS amplitudes appeared in the early AdS/CFT literature;
see \cite{Freedman:1998tz,Freedman:1998bj,Liu:1998ty,Arutyunov:2000py,Dolan:2004mu} for a sample of early papers and \cite{DHoker:2002nbb} for a review. In more recent times explicit loop-level diagrams have also been computed, see, for example,
\cite{Penedones:2010ue,Fitzpatrick:2011dm, Aharony:2016dwx, Alday:2017xua, Aprile:2017bgs, Giombi:2017hpr,Yuan:2018qva, Bertan:2018khc,  Bertan:2018afl, Ghosh:2018bgd, Ponomarev:2019ofr, Carmi:2019ocp, Meltzer:2019nbs, Albayrak:2020bso, Costantino:2020vdu, Carmi:2021dsn, Heckelbacher:2022fbx, Banados:2022nhj}, and they are all in agreement with CFT expectations.
To a large extent, the community takes for granted that AdS amplitudes are CFT correlators. It is the purpose of this note to provide an explicit proof that this is the case to all orders in bulk perturbation theory. We will discuss in detail the case the external operators are scalars, but all steps have a straightforward generalisation to spinning operators. It would be interesting to spell out all technical details but we leave this for future work. 

In the next section, we summarize the constraints imposed by conformal invariance on CFT correlations functions. Then in section \ref{sec:AdS} we show that AdS amplitudes satisfy these contraints. In particular, this derivation shows that the constants and functions of cross-ratios that appear in CFT correlators are determined in terms of bulk data. 
In section \ref{sec: diffeo} we show how the constraints of conformal invariance emerge from bulk diffeomorpshisms and illustrate how our results for scalar correlators extend to spinning ones by considering the case of conserved currents.
We finish with a discussion of our results in section \ref{sec: conclusion}.

\section{CFT correlators} \label{sec: CFT}

We review in this section the constraints of conformal invariance 
on CFT correlation functions of primary operators. This is a topic with long history,
see \cite{Polyakov:1970xd, Osborn:1993cr,Costa:2011mg} for some of the original literature and \cite{Ginsparg:1988ui, DiFrancesco:1997nk,Rychkov:2016iqz,Osborn1} for reviews. 

Conformal transformations are diffeomorphisms that results in a Weyl transformation:
under $\vec{x}\rightarrow \vec{x}\pr$,
\begin{equation}
    \dd s^2\rightarrow \dd s^{\prime\,2} = \Omega^2(\vec{x})\dd s^2\,.
\end{equation}
We work with Euclidean signature and coordinates $x^\alpha$, $\alpha=1, \ldots, d$. We will also use a vector notation, $\vec{x}=\{x^\alpha\}$.
In flat space, where $\dd s^2 = \dd\vec{x}^{\,2}$, these are given by:
\begin{alignat}{2} \label{conf_1}
    \text{Poincare:}\hs &x'^\alpha = a^\alpha_{\ \beta}x^\beta+a^\alpha\,, ,\hs&&\Omega(\vec{x}) = 1\,,\\
    \text{Dilation:}\hs &x'^\alpha = \lambda x^\alpha\,,\hs&&\Omega(\vec{x}) = \lambda\,,\\
    \text{Inversion:}\hs &x'^\alpha = \frac{x^\alpha} {\vec{x}^{\,2}}\,,\hs&&\Omega(\vec{x}) = \frac{1}{\vec{x}^{\,2}}\,, \label{inv}\\
    \text{Special conformal:}\hs &x'^\alpha = \frac{x^\alpha+b^\alpha\vec{x}^{\,2}}{1+2\vec{b}\cdot\vec{x}+\vec{b}^{\,2}\vec{x}^{\,2}}\,,\hs&&\Omega(\vec{x}) = \frac{1}{1+2\vec{b}\cdot\vec{x}+\vec{b}^{\,2}\vec{x}^{\,2}}\,. \label{conf_4}
\end{alignat}
The factors $\Omega(\vec{x})$ are related to the Jacobian $|\partial\vec{x}\pr/\partial\vec{x}|$ via
$\Omega(\vec{x})=|\partial \vec{x}'/\partial \vec{x}|^{1/d}$ and $\partial x'^\alpha/\partial x^\beta = \Omega(\vec{x})R^\alpha_{\ \beta}(\vec{x})$, where $R^\alpha_{\ \beta}\in O(d)$ is the orthogonal matrix:
 \begin{equation} \label{eq:R}
 R^\alpha_{\ \beta}(\vec{x}) = \left\{a^\alpha_{\ \beta},\ \delta^\alpha_\beta,\ I^\alpha_\beta(\vec{x}),\ I^\alpha_\gamma\left(\frac{\vec{x}}{\vec{x}^{\,2}}+\vec{b}\right)I^\gamma_\beta(\vec{x})\right\}\,,
 \end{equation}
for Poincare, dilations, inversions and special conformal transformation, correspondingly, and $\det R=\pm 1$ with $-1$ for inversions, $+1$ for the rest, where 
\begin{equation} \label{eq: I_def}
   I^\alpha_\beta(\vec{x})=\delta^\alpha_\beta-2 \frac{x^\alpha x_\beta}{\vec{x}^{\,2}}\, .  
\end{equation}
One may check conformal transformations satisfy,
\begin{equation}
    (\vec{x}\pr_1-\vec{x}\pr_2)^2 = \Omega(\vec{x}_1)\Omega(\vec{x}_2)(\vec{x}_1-\vec{x}_2)^2\,.
\end{equation}
and
\begin{equation}
    I_{\alpha\beta}(\vec{x}_1\pr-\vec{x}_2\pr) = R_\alpha^{\ \gamma}(\vec{x}_1)R_\beta^{\ \delta}(\vec{x}_2)I_{\gamma\delta}(\vec{x}_1-\vec{x}_2)\,.
\end{equation}

For concreteness, and to keep the technicalities to the minimum we will primarily focus on scalar operators, and we will briefly discuss spinning operator at the end of this section.

Scalar primary operators $\O$ of dimension $\Delta$ transform as 
\begin{equation}
\O'(\vec{x}\pr) = \Omega(\vec{x})^{-\Delta} \O(\vec{x}), \qquad
\end{equation}
and $n$-point functions should therefore satisfy
\begin{align}   
\langle\mathcal{O}_1(\vec{x}_1\pr)\dotsm\mathcal{O}_n(\vec{x}_n\pr)\rangle &= \Omega(\vec{x}_1)^{-\Delta_1}\dotsm \Omega(\vec{x}_n)^{-\Delta_n}\langle\mathcal{O}_1(\vec{x}_1)\dotsm\mathcal{O}_n(\vec{x}_n)\rangle\,, \label{npt_scalar} 
\end{align}

Following the presentation in \cite{Bzowski:2020kfw}, the solution of \eqref{npt_scalar} is given by
\begin{equation} \label{CFTn}
\langle \O_1 (\vec{x}_1) \ldots \O_n(\vec{x}_n) \rangle = \frac{C_n(u_{ijkl})}{\displaystyle \prod_{1 \leq i < j \leq n} (x^2_{ij})^{\Delta^{(n)}_{ij}}},
\end{equation}
where $ x^2_{ij}\equiv |\vec{x}_{ij}|^2\equiv (\vec{x}_i-\vec{x}_j)^2$ and
the parameters $\Delta^{(n)}_{ij}$ are related to the scaling dimensions by the relations,
\begin{equation} \label{deltaij}
\Delta_i =  \sum_{j=1}^n \Delta^{(n)}_{ij}, \qquad i=1,2,\ldots,n\, ,
\end{equation}
where we have assumed without loss of generality that 
$\Delta^{(n)}_{ji} = \Delta^{(n)}_{ij}, \Delta^{(n)}_{ii} = 0$.

The functions $C_n(u_{ijkl})$ are arbitrary functions of the conformal cross ratios,
\begin{equation}
u_{ijkl} = \frac{x_{ij}^2 x_{kl}^2}{x_{ik}^2 x_{jl}^2}, \qquad i \neq j \neq k \neq l,\, .
\end{equation}
These functions encode theory-specific information. Cross-ratios exist from 4-point function on, so $C_2$ and $C_3$ are constants. Not all cross ratios are independent. For instance:
\begin{equation}
    u_{ijkl} = u_{jilk} = u_{klij} = u_{lkji} = \frac{1}{u_{ikjl}} = \frac{1}{u_{jlik}} = \frac{1}{u_{kilj}} = \frac{1}{u_{ljki}}\, ,
\end{equation}
and there are more relations involving product of cross ratios. A simple counting suggests there are $n(n-3)/2$ independent cross-ratios (this is an over-counting when $n>d+2$, see for example \cite{Osborn1} -- this is not going to play a role here). One may choose the following combinations as independent 
cross-ratios,
\begin{equation} \label{cross-ratios}
    u_i = u_{123i} = \frac{x_{12}^2x_{3i}^2}{x_{13}^2x_{2i}^2}\,,\hs v_i = u_{321i} = \frac{x_{1i}^2x_{23}^2}{x_{13}^2x_{2i}^2}\,,\hs w_{ij} = u_{23ij} = \frac{x_{23}^2x_{ij}^2}{x_{2i}^2x_{3j}^2}\,,
\end{equation}
where $i,j=4,\dotsc,n$ and $i<j$.

Equation \eqref{deltaij} is a set of $n$ linear equations that may be used to determine $\Delta^{(n)}_{ij}$ 
given $\Delta_i$. When $n=2$, we find 
\begin{equation}
 \Delta_1=\Delta^{(2)}_{12}=\Delta_2,   
\end{equation}
 encoding the fact that only operators with same dimension have non-vanishing 2-point functions. When $n=3$, the unique solution is
 \begin{equation}
     \Delta_{ij}^{(3)} = \Delta_i+\Delta_j-\frac{\Delta_T}{2}, %\qquad \Delta_T = \sum_{i=1}^3 \Delta_i
 \end{equation}
where $\Delta_T$ denotes the sum over all dimensions, $\Delta_T = \sum \Delta_i$. 

For $n>3$ there are more unknowns than equations: $\Delta_{ij}^{(n)}$ is a symmetric hollow matrix ({\it i.e.} symmetric with zero in the diagonals)  so it has 
$n(n-1)/2$ independent matrix elements and we have $n$ equations to satisfy. It follows that the solution of \eqref{deltaij} is determined up to $n (n-3)/2$ constants, which is precisely the number of cross-ratios. The general solution is given by
\begin{equation}
   \Delta_{ij}^{(n)} = \hat{\Delta}_{ij}^{(n)} + \delta^{(n)}_{ij} 
\end{equation}
where
\begin{equation} \label{delta_ref}
    \hat{\Delta}_{ij}^{(n)} =  \frac{1}{n-2}\left(\Delta_i+\Delta_j-\frac{\Delta_T}{n-1}\right), \qquad i<j \, .
\end{equation}
is a particular solution and $\delta^{(n)}_{ij}$ is a symmetric hollow matrix satisfying the homogeneous linear equations:  
\begin{equation}
\sum_{j=1}^n \delta^{(n)}_{ij} =0, \qquad i=1,2,\ldots,n. 
\end{equation}
These equations may be solved by linearly expressing any $n$ of the $n(n-1)/2$ parameters $\delta^{(n)}_{ij} $ in terms of the remaining $n(n-1)/2 - n = n(n-3)/2$ ones. For example, when $n=4$ we  may solve $\delta^{(4)}_{12}, \delta^{(4)}_{23}, \delta^{(4)}_{24}, \delta^{(4)}_{34}$ in terms of 
$\delta^{(4)}_{13}$ and $\delta^{(4)}_{14}$:
\begin{equation}
    \delta^{(4)}_{12}=\delta^{(4)}_{34} = - \delta^{(4)}_{13} - \delta^{(4)}_{14}, \quad \delta^{(4)}_{23}=\delta^{(4)}_{14}, 
    \quad \delta^{(4)}_{24}=\delta^{(4)}_{13}\, .
\end{equation}
Then 
\begin{equation}
    \langle \O_1 (\vec{y}_1) \ldots \O_4(\vec{y}_4) \rangle = \frac{C_4(u_4, v_4)}{\displaystyle \prod_{1 \leq i < j \leq 4} (x^2_{ij})^{\Delta^{(4)}_{ij}}}
    = \frac{\hat{C}_4(u_4, v_4)}{\displaystyle \prod_{1 \leq i < j \leq 4} (x^2_{ij})^{\hat{\Delta}^{(4)}_{ij}}},
\end{equation}
where $\hat{C}_4(u_4, v_4) = C_4(u_4, v_4) u_4^{\delta^{(4)}_{13} + \delta^{(4)}_{14}} v_4^{-\delta^{(4)}_{14}}$. Thus the freedom in the solution of \eqref{deltaij}
just amounts to redefining the arbitrary function of cross-ratios. The same is true for any $n$. To have an unambiguous definition of the function of cross-ratios one needs to choose a solution of \eqref{deltaij}.

The formulas for spinning operators are similar but more involved. Here we will quote the results for the case of vector primaries as we will needed it later. Vector primaries $\J_\alpha$ of dimension $\Delta$ transform
\begin{equation} \label{eq: vec_tr}
    \J'_\alpha(\vec{x}\pr) = \Omega^{-\Delta}(\vec{x})R_\alpha^{\ \beta}(\vec{x})\J_\beta(\vec{x})\,,
\end{equation}
and this implies that $n$-point function should satisfy,
\begin{equation} \label{eq:npt_vec}
 \langle\mathcal{J}^1_{\alpha_1}(\vec{x}_1\pr)\dotsm\mathcal{J}^n_{\alpha_n}(\vec{x}_n\pr)\rangle = \Omega(\vec{x}_1)^{-\Delta_1}\dotsm \Omega(\vec{x}_n)^{-\Delta_n}
R_{\alpha_1}^{\ \beta_1}(\vec{x}_1) \cdots R_{\alpha_n}^{\ \beta_n}(\vec{x}_n)
\langle\mathcal{J}^1_{\beta_1}(\vec{x}_1)\dotsm\mathcal{J}^n_{\beta_n}(\vec{x}_n)\rangle\,,
\end{equation}
where $R_\alpha^{\ \beta}(\vec{x})$ is given in \eqref{eq:R}.

\section{AdS amplitudes} \label{sec:AdS}

The objects of interest are AdS amplitudes, which may be computed via Witten diagrams. 
The basic structure is well known \cite{Witten}: a Witten diagram  for an $n$-point function is constructed by $n$ bulk-to-boundary propagators which are linked to a number of bulk-to-bulk propagators connected via bulk vertices, which are integrated over all of AdS. 

As just reviewed, conformal invariance fixes the form of 2-point and 3-point functions, up to a number of constants, and the form of higher-point functions up to a functions of cross-ratios. The constants and the function of cross-ratios depend on the specific CFT but the form of the correlators is independent of it. In AdS/CFT correspondence the AdS isometries play the role of conformal transformations, so one should be able to establish the same results using AdS isometries only. We will show that this is indeed the case, and along the way we will also show  the relation of the arbitrary constants and functions of cross-ratios with bulk quantities.  We will establish this result to all orders in bulk perturbation theory and for scalar correlators. We will discuss the generalisation to general spinning operators afterwards.

\subsection{AdS isometries as constrained conformal transformations} \label{sec:AdSIso}

We work in Euclidean signature and use coordinates where AdS metric is given by
\begin{equation}
\dd s^2 = g_{\mu \nu} \dd x^\mu \dd x^\nu= \frac{\ell^2}{z^2} (\dd z^2 + \dd \vec{x}^{\,2}) = \ell^2\frac{\delta_{\mu \nu} \dd x^\mu \dd x^\nu}{z^2},  
\end{equation}
where $\ell$ is the AdS radius (set to 1 in the paper). The conformal boundary is at $z=0$ and this is the place where we need to impose boundary conditions. We will denote bulk point by $x^\mu=(z, x^\alpha)=(z, \vec{x})$, where $\mu=0, 1, \ldots, d$ is a bulk index, $\alpha=1, 2 , \ldots, d$ is a boundary index and $x^0=z$ is the radial coordinate. 

It is well known that the AdS metric is invariant under the following transformations (AdS isometries),
\begin{alignat}{3}
    &z' = z\,,\hs&&x'^\alpha = a^\alpha_{\ \beta}x^\beta+a^\alpha\ &&\rightarrow \text{Poincare}\,,\label{is1}\\
    &z'=\lambda z\,,\hs&&x'^\alpha = \lambda x^\alpha\ &&\rightarrow \text{Dilation}\,,\\
    &z'=\frac{z}{(z^2+ \vec{x}^{\,2})}\,,\hs&&x'^\alpha = \frac{x^\alpha}{(z^2+ \vec{x}^{\,2})}\ &&\rightarrow \text{Inversion}\,,\\
    &z'=\frac{z}{1+2\vec{b}\cdot\vec{x}+\vec{b}^{\,2}(z^2+ \vec{x}^{\,2})}\,,\hs&&x'^\alpha = \frac{x^\alpha+b^\alpha (z^2+ \vec{x}^{\,2})}{1+2\vec{b}\cdot\vec{x}+\vec{b}^{\,2}(z^2+ \vec{x}^{\,2})}\hs &&\rightarrow \text{Special conformal}\label{is4}
\end{alignat}
where we have also indicated the conformal transformation they limit to at the conformal boundary as $z \to 0$. 

It is less known that these transformations can be thought of as constrained flat-space $(d+1)$-dimensional conformal transformations. We will denote these transformations as in \eqref{conf_1}-\eqref{conf_4} but with the parameters carrying a tilde (and the indices being $(d+1$-dimensional indices): $a^\alpha{}_\beta \to \tilde{a}^\mu{}_\nu, a^\alpha \to \tilde{a}^\mu, \lambda \to \tilde{\lambda}, b^\alpha \to \tilde{b}^\mu$. Under such transformations
\begin{equation}
   \delta_{\mu \nu} \dd x^{\prime \mu}  \dd x^{\prime \nu}  = \tilde{\Omega}(x)^2 \delta_{\mu \nu} \dd x^\mu \dd x^\nu,
\end{equation}
where $\tilde{\Omega}(x)$ is $(d+1)$ version of $\Omega(\vec{x})$ in \eqref{conf_1}-\eqref{conf_4}.
For these conformal transformations to be AdS isometries the transformation of $z$ must cancel the factor of $\tilde{\Omega}$:
\begin{equation} \label{z_trans}
    z' = \tilde{\Omega}(x) z\, .
\end{equation}
This is indeed satisfied if we impose:
\begin{equation}
    \tilde{a}^z_{\ \nu}=\delta^z_\nu, \qquad \tilde{a}^\nu_{\ z}=\delta^\nu_z, \qquad \tilde{a}^z = 0, \qquad \tilde{b}^z=0\, .
\end{equation}
Thus, altogether and after dropping the tildes we obtain
\begin{alignat}{2} \label{AdSconf_1}
    x'^\mu &= a^\mu_{\ \nu}x^\nu+a^\mu\,, \qquad {\rm with}\ a^z_{\ \nu}=\delta^z_\nu, \quad a^\mu_{\ z} =\delta^\mu_z,\quad   a^z = 0 \,,\\
    x'^\mu &= \lambda x^\mu\, ,\\
    x'^\mu &= \frac{x^\mu} {x^{\,2}}\,,\, \label{invads}\\
    x'^\mu &= \frac{x^\mu+b^\mu x^{\,2}}{1+2 b\cdot x+ b^{\,2} x^{\,2}}\,, \qquad {\rm with}\ b^z=0\,, \label{AdSconf_4}
\end{alignat}
where $x^2=\delta_{\mu \nu} x^\mu x^\nu = z^2 + \vec{x}^{\,2}$, $b \cdot x = \delta_{\mu \nu} b^\mu x^\nu = \vec{b}\cdot\vec{x}$. One may readily check that
\eqref{AdSconf_1}-\eqref{AdSconf_4} agree with \eqref{is1}-\eqref{is4}. 

The advantage of viewing AdS isometries as constrained conformal transformations is that we can immediately inherit all CFT properties that are independent of specific rotations $a^\mu_{\ \nu}$ or translations $a^\mu,b^\mu$. For instance, the Jacobian of AdS isometries can be immediately obtained:
\begin{equation}\label{adsjacob}
    \frac{\partial x'^\mu}{\partial x^\nu} = \tilde{\Omega}(x)\tilde{R}^\mu_{\ \nu}(x)\,,
\end{equation}
where $\tilde{\Omega}$ (the one from $z'=\tilde{\Omega}z$) and $\tilde{R}^\mu_{\ \nu}\in O(d+1)$ are those of a CFT for constrained rotations and translations:
\begin{alignat}{2}
    \text{Constrained Poincare:}\hs&\tilde{\Omega}(x) = 1\,,\hs&&\tilde{R}^\mu_{\ \nu}(x) = a^\mu_{\ \nu}\,,\hs a^z_{\ \nu}=\delta^z_\nu\,,\ a^\mu_{\ z}=\delta^\mu_z\,,\\
    \text{Dilation:}\hs &\tilde{\Omega}(x)=\lambda\,,\hs&&\tilde{R}^\mu_{\ \nu}(x) = \delta^\mu_\nu\,,\\
    \text{Inversion:}\hs&\tilde{\Omega}(x)=\frac{1}{x^2}\,,\hs&&\tilde{R}^\mu_{\ \nu}(x) = I^\mu_\nu(x) = \delta^\mu_\nu-2\frac{x^\mu x_\nu}{x^2}\,,\\
    \text{Constrained SCT:}\hs&\tilde{\Omega}(x)=\frac{1}{1+2\vec{b}\cdot\vec{x}+\vec{b}^{\,2}x^2}\,,\hs&&\tilde{R}^\mu_{\ \nu}(x) = I^\mu_\rho\left(\frac{x}{x^2}+\vec{b}\right)I^\rho_\nu(x)\,.
\end{alignat}
This implies for example that the inversion property of AdS is inherited from that of flat space:
\begin{equation}\label{tr1}
    (x_1'-x_2')^2 = \tilde{\Omega}(x_1)\tilde{\Omega}(x_2)(x_1-x_2)^2\,,
\end{equation}
and
\begin{equation}\label{tr2}
    I_{\mu\nu}(x_1'-x_2') = \tilde{R}_\mu^{\ \rho}(x_1)\tilde{R}_\nu^{\ \sigma}(x_2)I_{\rho\sigma}(x_1-x_2)\,.
\end{equation}

We can further obtain useful formulas by taking the limit of bulk points to the boundary. In this limit, $\tilde{\Omega}$ and the boundary components of $\tilde{R}^\mu_{\ \nu}$ reduce to those of an unconstrained CFT in $d$ dimensions:
\begin{equation}
    \lim_{z\rightarrow0}\ \tilde{\Omega}(x) = \Omega(\vec{x})\,,\hs\lim_{z\rightarrow0}\ \tilde{R}^\alpha_{\ \beta}(x) = R^\alpha_{\ \beta}(\vec{x})\,,
\end{equation}
and thus one also recovers the Jacobian:
\begin{equation} \label{Jac_limit}
    \lim_{z\rightarrow0}\ \frac{\partial x'^\alpha}{\partial x^\beta} = \lim_{z\rightarrow0}\ \tilde{\Omega}(x) \tilde{R}^\alpha_{\ \beta}(x) = \Omega(\vec{x}) R^\alpha_{\ \beta}(\vec{x})\,.
\end{equation}
When one of the points in \eqref{tr1} and \eqref{tr2} are taken to the boundary, one obtains the useful relations:
\begin{equation}\label{tr1bdry}
    (x'-\vec{y}\pr)^2 = \tilde{\Omega}(x)\Omega(\vec{y})(x-\vec{y})^2\,,
\end{equation}
and:
\begin{equation}\label{tr2bdry}
    I_{\mu \alpha}(x'-\vec{y}\pr) = \tilde{R}_\mu^{\ \nu}(x)R_\alpha^{\ \beta}(\vec{y})I_{\nu \beta}(x-\vec{y})\,.
\end{equation}

\subsection{AdS propagators}

The bulk-to-boundary propagator for a bulk field dual to an operator of dimension $\Delta$ is the regular solution of the bulk equation
\begin{equation}
    (-\Box_g+m^2) K_\Delta(x_1,\vec{x}_2) =0
\end{equation}
where $m^2=\Delta (\Delta -d)$ and it is given by
\begin{equation}\label{eq:bulk bound prop}
K_\Delta(x_1, \Vec{x}_2) =c_\Delta \left( \frac{z_1}{(x_1 -\vec{x}_2)^2}\right)^\Delta , \qquad c_\Delta=\frac{\Gamma(\Delta)}{\pi^{\frac{d}{2}}\Gamma\left(\Delta-\frac{d}{2}\right)}.
\end{equation}
where $(x_1-\vec{x}_2)^2=z_1^2+(\vec{x}_1-\vec{x}_2)^2$. This is normalized such that as we approach the AdS boundary the propagator tends to a delta function, 
\begin{align}
   \lim_{z_1 \to 0} K_{\Delta}(x_1,\vec{x}_2)\rightarrow z_1^{d-\Delta}\delta(\vec{x}_1-\vec{x}_2)\, .
\end{align}
The bulk-to-boundary propagator $K_\Delta(x_1, \Vec{x}_2)$  transforms as a CFT primary field of dimension $\Delta$ at $\vec{x}_2$ under the AdS isometries \eqref{is1}-\eqref{is4} acting simultaneously on $x_1$ and $\Vec{x}_2$:
\begin{equation} \label{K_tranf}
 K_\Delta(x_1', \Vec{x}_2\pr) = \Omega(\vec{x}_2)^{-\Delta} K_\Delta(x_1, \Vec{x}_2)\, ,  
\end{equation}
where the factors of $\Omega$ are those given in \eqref{conf_1}-\eqref{conf_4}.
This is most easily shown using the perspective of the AdS isometries as constrained conformal transformations. Indeed, using \eqref{z_trans} and \eqref{tr1bdry} we obtain
\begin{equation}
    K_\Delta(x_1', \Vec{x}_2\pr) =c_\Delta \left( \frac{z'_1}{(x'_1 -\vec{x}\pr_2)^2}\right)^\Delta
    = c_\Delta \left( \frac{\tilde{\Omega}(x_1) z_1}{\tilde{\Omega}(x_1) \Omega(\vec{x}_2)(x_1 -\vec{x}_2)^2}\right)^\Delta = \Omega(\vec{x}_2)^{-\Delta} K_\Delta(x_1, \Vec{x}_2)
\end{equation}
We will also explain in section \ref{sec: diffeo} that this transformation rule follows from bulk diffeomorphism invariance.

The bulk-to-bulk propagator for the same field is the regular solution of
the equation 
\begin{equation}
    (-\Box_g+m^2)G_\Delta(x_1,x_2) = \frac{1}{\sqrt{g}}\delta(x_1-x_2)\, ,
\end{equation}
with normalizable behavior at infinity, $G_\Delta \sim z_1^\Delta$ as $x_1$ approaches the conformal boundary (and $\sim z_2^\Delta$ when $x_2$ approaches the conformal boundary). AdS invariance implies that the propagator is a function of an AdS invariant distance, which we may take to be the chordal distance,
\begin{equation} \label{eq:chordal}
    \xi=\frac{2z_1 z_2}{z_1^2+z_2^{2}+(\Vec{x}_1-\Vec{x}_2)^2}\, .
\end{equation}
The invariance of the chordal distance under transformations \eqref{is1}-\eqref{is4} (or equivalently \eqref{AdSconf_1}-\eqref{AdSconf_4}) that act simultaneously on both $x_1$ and $x_2$ follows by inspection upon use of \eqref{z_trans} and \eqref{tr1}. By explicit computation,
\begin{equation}\label{eq:bulk bulk prop}
G_\Delta(x_1, x_2)=\frac{2^{-\Delta}c_\Delta}{2\Delta-d}\xi^\Delta\,_2 F_1\left(\frac{\Delta}{2}, \frac{\Delta+1}{2}, \Delta-\frac{d}{2}+1;\xi^2 \right)\, .
\end{equation}
It follows that 
\begin{equation}
    G_\Delta(x_1', x_2')=G_\Delta(x_1, x_2)\, .
\end{equation}

One may similarly obtain the transformation properties for propagators of spinning fields. We report here the result for the bulk-to-boundary propagator of an (Abelian) gauge field, as this is a case we discuss later. The bulk-to-boundary propagator has been obtained in the early AdS/CFT literature  \cite{Freedman1}. Up to gauge transformations it is given by
\begin{equation}
    K_{\mu \alpha}(x_1,\vec{x}_2) = C \frac{z_1^{d-2}}{[(x_1-\vec{x}_2)^2]^{d-1}} I_{\mu \alpha}(x_1-\vec{x}_2)\,,
\end{equation}
where $C$ is a constant and $I_{\mu \alpha}$ the inversion tensor, with $\mu$ and $\alpha$ bulk and boundary indices, respectively.  Using \eqref{z_trans}, \eqref{tr1bdry} and \eqref{tr2bdry} one may work out how this bulk-to-boundary propagator transforms under bulk isometries:
\begin{align} \label{Kvec_tra}
    K_{\mu \alpha}(x_1',\vec{x}\pr_2) &= C\frac{z_1'^{d-2}}{\left[(x_1'-\vec{x}\pr_2)^2\right]^{d-1}}I_{\mu\alpha}(x_1'-\vec{x}_2\pr) \nonumber \\
    &= C\frac{\tilde{\Omega}^{d-2}(x_1)z_1^{d-2}}{\left[\tilde{\Omega}(x_1)\Omega(\vec{x}_2)(x_1-\vec{x}_2)^2\right]^{d-1}}\tilde{R}_\mu^{\ \nu}(x_1)R_\alpha^{\ \beta}(\vec{x}_2)I_{\nu \beta}(x_1-\vec{x}_2)\,, \nonumber\\
    &= \Omega^{-(d-1)}(\vec{x}_2)\tilde{\Omega}^{-1}(x_1)\tilde{R}_\mu^{\ \nu}(x_1)C\frac{z_1^{d-2}}{\left[(x_1-\vec{x}_2)^2\right]^{d-1}}I_{\nu \beta}(x_1-\vec{x}_2)R_\alpha^{\ \beta}(\vec{x}_2)\,, \nonumber \\
    &= \Omega^{-(d-1)}(\vec{x}_2)\frac{\partial x_1^\nu}{\partial x_1'^\mu}K_{\nu \beta}(x_1,\vec{x}_2)R_\alpha^{\ \beta}(\vec{x}_2)\,.
\end{align}
It follows that the vector bulk-to-boundary propagators transforms as a vector in the bulk index $\mu$ and a CFT conserved current in the boundary index $\alpha$ (compare with \eqref{eq: vec_tr} with $\Delta=d-1$). We will rederive this transformation property from bulk diffeomorphism invariance in section \ref{sec: diffeo}.
 
\subsection{AdS amplitudes are CFT correlators}

We now discuss the computation of AdS amplitudes, {i.e.} bulk $n$-point functions with all legs in AdS boundary. This can be computed via Witten-Feynman diagrams, involving $n$ bulk-to-boundary propagators connecting the $n$ boundary points to an ``amputated'' bulk $n$-point function $G_n(x_1, \ldots, x_n)$ and integrating over $x_1, \ldots, x_n$. The amputated bulk $n$-point function is constructed from bulk-to-bulk propagator connected via vertices that come from the bulk action, and integrating over the position of each vertex. As long as the bulk action is invariant under AdS isometries, the invariance of the bulk-to-bulk propagator guarantees that $G_n(x_1, \ldots, x_n)$ is also invariant under \eqref{is1}-\eqref{is4} that act simultaneously on all $x_1, \ldots, x_n$,
\begin{equation} \label{scr_inv}
 G_n(x'_1,\dotsc,x'_n)=G_n(x_1,\dotsc,x_n)\, .   
\end{equation}
This could have been invalidated by short-distance singularities, but as we discuss in section \ref{sec:UVreg} we can regulate the short-distance singularities while respecting the AdS isometries. More generally, \eqref{scr_inv} is guaranteed by diffeomorphism invariance (in a theory with no diffeomorphism anomalies), and we will discuss in section \ref{sec: diffeo} the extension to tensorial correlators. When the bulk points $x_i$ tend to the boundary, IR divergences appear. These correspond via the AdS/CFT correspondence to UV divergences in the dual CFT and lead to conformal anomalies and anomalous dimensions. This will be discussed in section \ref{sec:UVreg}, but for ease in presentation we suppress the IR issues in this subsection.  We will now show that the dependence of the correlators on the external positions $\Vec{y}_i$ is the same with that of a CFT, without computing any integral.
  
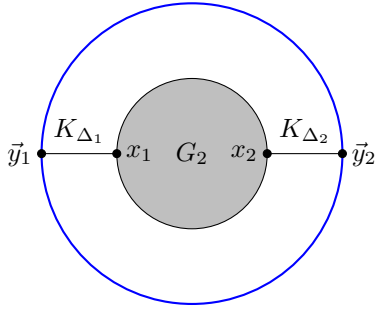
\begin{figure}[h]
    \centering
    \begin{tikzpicture}
        \draw[blue,thick] (0,0) circle (2cm);
        \filldraw[color=black,fill=black!25] (0,0) circle (1cm) node{$G_2$};
        \draw (-2,0) node[left]{$\vec{y}_1$} -- (-1,0) node[right]{$x_1$};
        \draw (1,0) node[left]{$x_2$} -- (2,0) node[right]{$\vec{y}_2$};
        \node [above] at (-1.5, 0.01) {${K}_{\Delta_1}$};
        \node [above] at (1.5, 0.01) {${K}_{\Delta_2}$};
        \filldraw[black] (-2,0) circle (1.5pt);
        \filldraw[black] (-1,0) circle (1.5pt);
        \filldraw[black] (1,0) circle (1.5pt);
        \filldraw[black] (2,0) circle (1.5pt);
    \end{tikzpicture}
    \caption{General 2-point function. The blue (outer) circle represents the boundary of AdS, and the shaded region represents general (loop) interactions that connect the bulk points $x_1, x_2$.} \label{fig:2pt}
\end{figure}

\paragraph{2-point function} Let us start with the 2-point function, whch is 
illustrated in Fig. \ref{fig:2pt},
\begin{equation}
    I_2(\vec{y}_1,\vec{y}_2) = \int_{x_1}\int_{x_2}K_{\Delta_1}(x_1,\vec{y}_1)G_2(x_1,x_2)K_{\Delta_2}(x_2,\vec{y}_2)\ ,
\end{equation}
where we use the shorthand notation,
$\int_x = \int d^{d+1} x \sqrt{\det g}$. 
We can extract all the dependence of the external points $\vec{y}_1$ and $\vec{y}_2$ from the integral by performing change variables in integration variables $x_1$ and $x_2$ 
that account to AdS isometries. First, note that by using the inversion property of the bulk-to-boundary propagators we find,
\begin{equation} \label{eq:inv}
    I_2(\Vec{y}_1^{\,\prime}, \Vec{y}_2^{\,\prime}) = (\Vec{y}_1^{\,2})^{\Delta_1} (\Vec{y}_2^{\,2})^{\Delta_2} I_2(\Vec{y}_1, \Vec{y}_2)
\end{equation}
We now shift the integration variables $\Vec{x}_1, \Vec{x}_2$ by $\Vec{y}_1$ to obtain,
\begin{equation}
I_2 = \int_{x_1}\int_{x_2}K_{\Delta_1}(x_1,\vec{0})G_2(x_1,x_2)K_{\Delta_2}(x_2,\vec{y}_{21})
\end{equation}
where $\vec{y}_{21} = (\Vec{y}_2-\Vec{y}_1)$. We can now change variables by rescaling $x_1, x_2$ by $|\vec{y}_{12}|$ and use %the homogeneity property 
\eqref{K_tranf} to find,
\begin{equation} \label{eq:2pt_2}
    I_2 = \frac{1}{|\vec{y}_{12}|^{\Delta_1+\Delta_2}}\int_{x_1}\int_{x_2}K_{\Delta_1}(x_1,\vec{0})G_2(x_1,x_2)K_{\Delta_2}(x_2,\hat{y}_{21})
\end{equation}
Thus,
\begin{equation} \label{2pt_3}
    I_2(\Vec{y}_1, \Vec{y}_2) = \frac{C_2}{|\vec{y}_{12}|^{\Delta_1+\Delta_2}},
\end{equation}
with $C_2$ equal to,
\begin{equation} \label{c2}
    C_2 = \int_{x_1}\int_{x_2}K_{\Delta_1}(x_1,\vec{0})G_2(x_1,x_2)K_{\Delta_2}(x_2,\hat{y}_{21})
\end{equation}
Finally, rotational invariance implies that the integral does not depend on the direction specified by $\hat{y}_{21}$ and thus it is a constant.
Equation \eqref{2pt_3} should  be consistent with the transformation in \eqref{eq:inv} and this implies $\Delta_1=\Delta_2 =\Delta$, thus reproducing the expected CFT answer,
\begin{equation}
    I_2(\Vec{y}_1, \Vec{y}_2) = \frac{C_2}{|\vec{y}_{12}|^{2 \Delta}}.
\end{equation}

\begin{figure}[h]
    \centering
    \begin{tikzpicture}
        \draw[blue,thick] (0,0) circle (2cm);
        \filldraw[color=black,fill=black!25] (0,0) circle (1cm) node{$G_3$};
        \node [right] at (-1.5, 1.0) {${K}_{\Delta_1}$};
        \draw (-1.73,1) node[left]{$\vec{y}_1$} -- (-0.86,0.5) node[right]{$x_1$};
        \draw (0.86,0.5) node[left]{$x_2$} -- (1.73,1) node[right]{$\vec{y}_2$};
        \node [right] at (0.55, 1.0) {${K}_{\Delta_2}$};
        \draw (0,-1) node[above]{$x_3$} -- (0,-2) node[below]{$\vec{y}_3$};
        \node [right] at (0.01, -1.5) {${K}_{\Delta_3}$};
        \filldraw[black] (-1.73,1) circle (1.5pt);
        \filldraw[black] (-0.86,0.5) circle (1.5pt);
        \filldraw[black] (1.73,1) circle (1.5pt);
        \filldraw[black] (0,-1) circle (1.5pt);
        \filldraw[black] (0,-2) circle (1.5pt);
        \filldraw[black] (0.86,0.5) circle (1.5pt);
    \end{tikzpicture}
    \caption{General 3-point function. The blue (outer) circle represents the boundary of AdS, and the shaded region represents general (loop) interactions that connect the bulk points $x_1, x_2, x_3$. \label{fig:3pt}}
\end{figure}
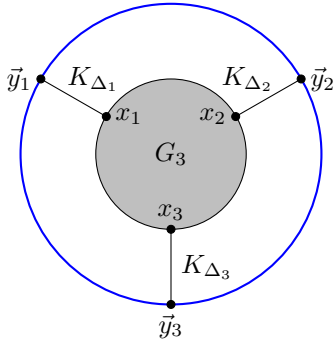

\paragraph{3-point function}  The general 3-point function, see Fig. \ref{fig:3pt}, is given by
\begin{equation}
    I_3(\vec{y}_1,\vec{y}_2,\vec{y}_3) = \int_{x_1}\int_{x_2}\int_{x_3} K_{\Delta_1}(x_1,\vec{y}_1)K_{\Delta_2}(x_2,\vec{y}_2)K_{\Delta_3}(x_3,\vec{y}_3)G_3(x_1,x_2,x_3),
\end{equation}
where $G_3(x_1, x_2, x_3)$ is the amputated bulk 3-point function.
We first shift the integration variables $\vec{x}_2, \vec{x}_2, \vec{x}_3$ by $\Vec{y}_1$ to obtain
\begin{equation}
    I_3 = \int_{x_1}\int_{x_2}\int_{x_3} K_{\Delta_1}(x_1,\vec{0})K_{\Delta_2}(x_2,\vec{y}_{21})K_{\Delta_3}(x_3,\vec{y}_{31})G_3(x_1,x_2,x_3)
\end{equation} 
Then we make a change of variable that amounts to an inversion on all integration variables and use the transformation of the bulk-to-boundary propagator \eqref{K_tranf} to obtain
\begin{equation}
I_3  = |\vec{y}^{\,\prime}_{12}|^{2\Delta_2}|\vec{y}^{\,\prime}_{13}|^{2\Delta_3}\int_{x_1}\int_{x_2}\int_{x_3}c_{\Delta_1} z_1^{\Delta_1}K_{\Delta_2}(x_2,\vec{y}^{\,\prime}_{21})K_{\Delta_3}(x_3,\vec{y}^{\,\prime}_{31}) G_3(x_1,x_2,x_3),    
\end{equation}
where here and in the remainder of this section prime indicates a (boundary) inversion.
\begin{equation} \label{eq:bdry_inv}
    \vec{y}^{\,\prime}=\frac{\vec{y}}{\vec{y}^2}\,.
\end{equation}
After this step, only two bulk-to-boundary propagators depend on the external positions, so we can proceed analogously to the case of 2-point function to obtain
\begin{equation}    I_3=\frac{|\vec{y}^{\,\prime}_{12}|^{2\Delta_2}|\vec{y}^{\,\prime}_{13}|^{2\Delta_3}}{|\vec{y}^{\,\prime}_{31}-\vec{y}^{\,\prime}_{21}|^{\Delta_2+\Delta_3-\Delta_1}}\int_{x_1}\int_{x_2}\int_{x_3}c_{\Delta_1} z_1^{\Delta_1}K_{\Delta_2}(x_2,\vec{0})K_{\Delta_3}(x_3,\hat{y}'_{31,21}) G_3(x_1,x_2,x_3)
\end{equation}
where $\hat{y}'_{31,21}$ is the unit vector of $\vec{y}^{\,\prime}_{31}-\vec{y}^{\,\prime}_{21}$, {\it i.e.}
\begin{equation}
    \hat{y}'_{31,21} = \frac{\vec{y}^{\,\prime}_{31}-\vec{y}^{\,\prime}_{21}}{|\vec{y}^{\,\prime}_{31}-\vec{y}^{\,\prime}_{21}|}\, .
\end{equation}
Let 
\begin{equation} \label{c3}
    C_3=\int_{x_1}\int_{x_2}\int_{x_3}c_{\Delta_1} z_1^{\Delta_1}K_{\Delta_2}(x_2,\vec{0})K_{\Delta_3}(x_3,\hat{y}'_{31,21}) G_3(x_1,x_2,x_3)
\end{equation}
Rotation invariance implies that $C_3$ is independent of $\hat{y}'_{31,21}$ that thus it is a constant. Using \eqref{eq:bdry_inv} to re-express the answer in terms of the original insertion points we finally get 
\begin{equation}
    I_3(\vec{y}_1,\vec{y}_2,\vec{y}_3) = \frac{C_3}{|\vec{y}_{12}|^{\Delta_1+\Delta_2-\Delta_3}|\vec{y}_{13}|^{\Delta_1+\Delta_3-\Delta_2}|\vec{y}_{23}|^{\Delta_2+\Delta_3-\Delta_1}}\, ,
\end{equation}
which is precisely the expected form for a CFT 3-point function. 

\paragraph{4-point functions}
The  general 4-point function, see Fig. \ref{fig:4pt}, is given by
\begin{equation}
    I_4(\vec{y}_1,\vec{y}_2,\vec{y}_3,\vec{y}_4) = \int_{x_1} \int_{x_2}\int_{x_3}\int_{x_4}K_{\Delta_1}(x_1,\vec{y}_1)K_{\Delta_2}(x_2,\vec{y}_2)K_{\Delta_3}(x_3,\vec{y}_3)K_{\Delta_4}(x_4,\vec{y}_4)G_4(x_1,x_2,x_3,x_4),
\end{equation}
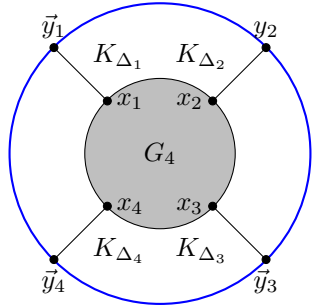
\begin{figure}[h]
    \centering
    \begin{tikzpicture}
        \draw[blue,thick] (0,0) circle (2cm);
        \filldraw[color=black,fill=black!25] (0,0) circle (1cm) node{$G_4$};
        \draw (-1.41,1.41) node[above]{$\vec{y}_1$} -- (-0.7,0.7) node[right]{$x_1$};
          \node [above] at (-0.55, 1.0) {${K}_{\Delta_1}$};
        \draw (0.7,0.7) node[left]{$x_2$} -- (1.41,1.41) node[above]{$y_2$};
        \node [above] at (0.55, 1.0) {${K}_{\Delta_2}$};
        \draw (0.7,-0.7) node[left]{$x_3$} -- (1.41,-1.41) node[below]{$\vec{y}_3$};
         \node [below] at (0.55, -1.0) {${K}_{\Delta_3}$};
        \draw (-0.7,-0.7) node[right]{$x_4$} -- (-1.41,-1.41) node[below]{$\vec{y}_4$};
         \node [below] at (-0.55, -1) {${K}_{\Delta_4}$};
        \filldraw[black] (-1.41,1.41) circle (1.5pt);
        \filldraw[black] (-0.7,0.7) circle (1.5pt);
        \filldraw[black] (0.7,0.7) circle (1.5pt);
        \filldraw[black] (1.41,1.41) circle (1.5pt);
        \filldraw[black] (0.7,-0.7) circle (1.5pt);
        \filldraw[black] (1.41,-1.41) circle (1.5pt);
        \filldraw[black] (-0.7,-0.7) circle (1.5pt);
        \filldraw[black] (-1.41,-1.41) circle (1.5pt);
    \end{tikzpicture}
    \caption{General 4-point function. The blue (outer) circle represents the boundary of AdS, and the shaded region represents general (loop) interactions that connect the bulk points $x_1, x_2, x_3, x_4$. \label{fig:4pt}}
\end{figure}
where $G_4(x_1,x_2,x_3,x_4)$ is the amputated bulk-to-bulk 4-point function. Following the same steps\footnote{In more detail: we translate the internal coordinates by $\vec{y}_1$, invert them together with the external coordinates, translate them again by $\vec{y}^{\ \prime}_{21}$, rescale them by $|\vec{y}^{\ \prime}_{31}-\vec{y}^{\ \prime}_{21}|$, and finally, write the inverted external points in terms of the original positions.} as in the case of 3-point functions we arrive at
\begin{align}
I_4 = & \frac{|\vec{y}_{12}|^{\Delta_3+\Delta_4-\Delta_1-\Delta_2}|\vec{y}_{13}|^{\Delta_2+\Delta_4-\Delta_1-\Delta_3}}{|\vec{y}_{14}|^{2\Delta_4}|\vec{y}_{23}|^{\Delta_2+\Delta_3+\Delta_4-\Delta_1}}  \nonumber \\
& \times \int_{x_1}\int_{x_2}\int_{x_3}\int_{x_4}c_{\Delta_1} z_1^{\Delta_1}K_{\Delta_2}(x_2,\vec{0})K_{\Delta_3}(x_3,\hat{y}'_{31,21})K_{\Delta_4}\left(x_4,\frac{|\vec{y}_{13}||\vec{y}_{24}|}{|\vec{y}_{14}||\vec{y}_{23}|}\hat{y}'_{41,21}\right)G_4(x_1,x_2,x_3,x_4)
\end{align}
where $\hat{y}'_{31,21}$ is the unit vector of $\vec{y}^{\,\prime}_{31}-\vec{y}^{\,\prime}_{21}$, 
$\hat{y}'_{41,21}$ the unit vector of $\vec{y}^{\,\prime}_{41}-\vec{y}^{\,\prime}_{21}$, and $\Vec{y}_{ij}^{\,\prime}=\Vec{y}_{ij}/\Vec{y}_{ij}^{\,2}$, is the inversion of 
$\Vec{y}_{ij}$ (and $\Vec{y}_{ij}=(\Vec{y}_i - \Vec{y}_j)$). Using the 
conformal cross-ratios from \eqref{cross-ratios} (and relabeling $u_4 \to u$ and $v_4 \to v$) we find that the 4-point function takes the expected form for a CFT 4-point function,
\begin{equation} \label{CFT4pt}
    I_4(\vec{y}_1,\vec{y}_2,\vec{y}_3,\vec{y}_4) = \frac{C_4(u,v)}{\displaystyle \prod_{1 \leq i < j \leq 4} (y^2_{ij})^{\Delta^{(4)}_{ij}}}\,,
\end{equation}
where the dimensions $\Delta_{ij}^{(4)}$ satisfy the conformal constraints \eqref{deltaij}, and
\begin{align} \label{eq:c4}
    C_4(u, v) =& u^{\Delta_{34}^{(4)}}v^{\Delta_{14}^{(4)}-\Delta_4}  \nonumber \\
& \times\int_{x_1}\int_{x_2}\int_{x_3}\int_{x_4}c_{\Delta_1} z_1^{\Delta_1}K_{\Delta_2}(x_2,\vec{0})K_{\Delta_3}(x_3,\hat{y}'_{31,21})K_{\Delta_4}\left(x_4,\frac{\hat{y}'_{41,21}}{\sqrt{v}}\right)G_4(x_1,x_2,x_3,x_4)
\end{align}
where in asserting that $C_4$ depends only on $u, v$ we used the fact that  rotational invariance implies that the integral may depend on $\hat{y}'_{31,21}$ and $\hat{y}'_{41,21}$ only via their inner product and as we now explain this inner product is a function of $u$ and $v$. Indeed, 
since  $\hat{y}'_{31,21}$ and $\hat{y}'_{41,21}$ are unit vectors their inner product depends only on the angle between them and conformal transformation preserves angles. $\hat{y}'_{31,21} \cdot \hat{y}'_{41,21}$ being conformal invariant that depends on four positions is necessarily is a function $u$ and $v$. We can compute this function explicit as follows. 
Reverting to the original variables we find,
\begin{equation}
    \hat{y}'_{31,21} \cdot \hat{y}'_{41,21} = \frac{y_{12}^2y_{13}y_{14}}{y_{23}y_{24}}\left(\frac{\vec{y}_{13}}{y_{13}^2}-\frac{\vec{y}_{12}}{y_{12}^2}\right)\cdot\left(\frac{\vec{y}_{14}}{y_{14}^2}-\frac{\vec{y}_{12}}{y_{12}^2}\right)\,.
\end{equation}
Expanding the product and using the formula
\begin{equation}
    \vec{y}_{1i}\cdot\vec{y}_{1j} = \frac{1}{2}\left(y_{1i}^2+y_{1j}^2-y_{ij}^2\right)\,,
\end{equation}
leads to the result
\begin{equation} \label{inner}
\hat{y}'_{31,21} \cdot \hat{y}'_{41,21} = \frac{1+v-u}{2\sqrt{v}}\,,
\end{equation}
which is a function of $u$ and $v$, as claimed. 

Note that the integral in \eqref{eq:c4} depends on $u$ only through the inner product in \eqref{inner}. This appears to be special to holographic CFT and it will be interesting to investigate its implications.

\paragraph{$n$-point function}
We now discuss the general case. Starting from
\begin{equation}
    I_n(\vec{y}_1,\dotsm,\vec{y}_n) = \int_{x_1}\dotsm\int_{x_n}\prod_{i=1}^n K_{\Delta_i}(x_i,\vec{y}_i)\ G_n(x_1,\dotsm,x_n)\,,
\end{equation}
\begin{figure}[h]
    \centering
    \begin{tikzpicture}
        \draw[blue,thick] (0,0) circle (2cm);
        \filldraw[color=black,fill=black!25] (0,0) circle (1cm) node{$G_n$};
        \draw (-1.41,1.41) -- (-0.7,0.7);
        \draw (-2,0) node[left]{$\vec{y}_1$} -- (-1,0) node[right]{$x_1$};
        \draw (2,0) node[right]{$\vec{y}_n$} -- (1,0) node[left]{$x_n$};
        \node [above] at (-1.5, 0) {${K}_{\Delta_1}$};
        \draw (0.7,0.7) -- (1.41,1.41);
        \node [above] at (1.5, 0) {${K}_{\Delta_n}$};
        \draw[loosely dotted,very thick] (-0.4,1.4) -- (0.4,1.4);
        \draw[loosely dotted,very thick] (-0.4,-1.4) -- (0.4,-1.4);
        \draw (0.7,-0.7) -- (1.41,-1.41);
        \draw (-0.7,-0.7) -- (-1.41,-1.41);
        \filldraw[black] (-1.41,1.41) circle (1.5pt);
        \filldraw[black] (-0.7,0.7) circle (1.5pt);
        \filldraw[black] (-2,0) circle (1.5pt);
        \filldraw[black] (-1,0) circle (1.5pt);
        \filldraw[black] (1,0) circle (1.5pt);
        \filldraw[black] (2,0) circle (1.5pt);
        \filldraw[black] (0.7,0.7) circle (1.5pt);
        \filldraw[black] (1.41,1.41) circle (1.5pt);
        \filldraw[black] (0.7,-0.7) circle (1.5pt);
        \filldraw[black] (1.41,-1.41) circle (1.5pt);
        \filldraw[black] (-0.7,-0.7) circle (1.5pt);
        \filldraw[black] (-1.41,-1.41) circle (1.5pt);
    \end{tikzpicture}
    \caption{General n-point function. The blue (outer) circle represents the boundary of AdS, and the shaded region represents general (loop) interactions that connect the bulk points $x_1,\dotsc,x_n$. \label{fig:npt}}
\end{figure}
which is represented by Fig. \ref{fig:npt}, and repeating the same steps one finds:
\begin{align}
    I_n = & \frac{y^{\Delta_T-2\Delta_1-2\Delta_2}_{12}y^{\Delta_T-2\Delta_1-2\Delta_3}_{13}}{y^{\Delta_T-2\Delta_1}_{23}\prod_{i=4}^ny^{2\Delta_i}_{1i}} \nonumber\\
    &\times \int_{x_1}\dotsm\int_{x_n}c_{\Delta_1}z_1^{\Delta_1}K_{\Delta_2}(x_2,\vec{0})K_{\Delta_3}(x_3,\hat{y}'_{31,21})\prod_{i=4}^n K_{\Delta_i}\left(x_i,\frac{y_{13}y_{2i}}{y_{1i}y_{23}}\hat{y}^\prime_{i1,21}\right)\ G_n(x_1 \ldots x_n)\,.
\end{align}
which may be processed to
\begin{equation}
    I_n(\vec{y}_1,\dotsc,\vec{y}_n) = \frac{C_n(u_i,v_i,w_{ij})}{\displaystyle \prod_{1 \leq i < j \leq n} (y^2_{ij})^{\Delta^{(n)}_{ij}}}\,,
\end{equation}
where the dimensions $\Delta_{ij}^{(n)}$ satisfy the conformal constraints \eqref{deltaij}, and
\begin{align} \label{cn}
    C_n = &\prod_{i=4}^nu_i^{\Delta_i-\Delta_{1i}^{(n)}-\Delta_{2i}^{(n)}-\underset{i<k\leq n}{\sum}\Delta_{ik}^{(n)}}v_i^{\Delta_{1i}^{(n)}-\Delta_i}\prod_{4\leq j<l\leq n}w_{jl}^{\Delta_{jl}^{(n)}}\nonumber\\
    &\times\int_{x_1}\dotsm\int_{x_n}c_{\Delta_1}z_1^{\Delta_1}K_{\Delta_2}(x_2,\vec{0})K_{\Delta_3}(x_3,\hat{y}'_{31,21})\prod_{i=4}^n K_{\Delta_i}\left(x_i,\frac{\hat{y}'_{i1,21}}{\sqrt{v_i}}\right)\ G_n(x_1,\dotsc,x_n)\,.
\end{align}
The remaining integral is a function of cross-ratios. Indeed, by rotational invariance it is a function of the inner product between the unit vectors $\hat{y}'_{i1,21}$. A similar computation as above leads to:
\begin{equation}
    \hat{y}'_{i1,21}\cdot\hat{y}'_{j1,21} = \frac{1}{2}\left(\sqrt{u_{1ij2}}+\sqrt{u_{1ji2}}-\sqrt{u_{12ij}}\sqrt{u_{12ji}}\right)\,.
\end{equation}
which is a function of cross-ratios, as claimed. For $i=j$, the product is just 1, as expected. For $j=3$ and $i>3$, it can be expressed in terms of $u_i$ and $v_i$:
\begin{equation}
    \hat{y}'_{i1,21}\cdot\hat{y}'_{31,21} = \frac{1+v_i-u_i}{2\sqrt{v_i}}\,,\hs i>3\,,
\end{equation}
while for $i,j>3$ also in terms of $w_{ij}$:
\begin{equation} \label{inner2}
    \hat{y}'_{i1,21}\cdot\hat{y}'_{j1,21} = \frac{v_i+v_j-u_jw_{ij}}{2\sqrt{v_iv_j}}\,,\hs i,j>3\,.
\end{equation}
As in the case of 4-point functions, the integral in \eqref{cn} appears to have special dependence on some of the cross-ratios ($u_i$ and $w_{ij}$) and it will be interesting to investigate the implications.

\subsection{Regularization and renormalization} \label{sec:UVreg}

We have just shown that AdS amplitudes can be brought to a form that manifestly satisfied the CFT Ward identities by a sequence of steps that involved changing integration variables amounting to AdS isometries. Such manipulations are well-posed if the integrals are finite. However, the integrals may diverge both in the UV and the IR. The UV divergences come from bulk loops, while the IR divergences are due to the infinite volume of AdS. The bulk IR divergences correspond to CFT UV divergences via the AdS/CFT correspondence.

One can regulate the UV divergences in a way that preserves the AdS 
invariance. This is expected as  bulk UV divergences are mapped by the AdS/CFT correspondence to IR divergences in the CFT. Such divergences should cancel on their own and they should not lead to breaking of the conformal symmetry.
The regulator amounts to separating (bulk) coincident points along a geodesic by affine distance $\tau$ \cite{Banados:2022nhj}, which thus acts an UV regulator. This results to modifying the argument 
of the bulk-to-bulk propagator by changing $\xi \to \xi/ \cosh \tau$ in \eqref{eq:bulk bulk prop}, recovering the prescription in \cite{Bertan:2018afl, Bertan:2018khc}. We will discuss in detail this regulator in \cite{upcoming}, where we will also show that it can be derived from a regulated action. Since the regulated theory is invariant under AdS isometries, all steps outlined above are valid in the regulated theory. The discussion in these papers is about bulk scalar propagators, but we expect the results to extend to general tensorial fields (the metric, gauge fields, antisymmetric tensor fields, etc.). 

The issues with the IR divergences is more subtle. These correspond to UV divergences in the dual CFT and such divergences give rise to anomalous dimensions and conformal anomalies, and thus the breaking of conformal symmetry is inevitable. One can regulate the IR divergence by imposing an explicit IR cut-off, $z \geq \varepsilon$, as in the original works on holographic renormalization \cite{Henningson:1998gx, deHaro:2000vlm, Kostas}. In this case the explicit cut-off results into additional terms when one follows the manipulations described above. This has already been discussed in  \cite{Banados:2022nhj} and we will discuss in detail how to compute such integrals in \cite{upcoming}. The results is that the AdS amplitudes still take the form of the CFT correlators but now the dimensions may renormalize (there are anomalous dimenions) and conformal anomalies appear.

An alternative approach is to use dimensional regularization, where the spacetime dimension $d$ and dimensions $\Delta_i$ of operators are shifted as in \cite{Bzowski_2014}. For generic values of  
$\Delta_i$ the correlators may be defined by analytic continuation. For such cases the analysis above holds unchanged. However, there are also cases where genuine singularities appear and boundary counterterms and renormalization is needed \cite{Bzowski:2015pba, Bzowski_2018A, Bzowski_2018B, Bzowski:2019kwd}. In the case where both UV and IR issues are present one would need to renormalize the parameters in the bulk action (masses and coupling constants), the fields that specify the boundary conditions (sources of the dual operators) and add appropriate boundary counterterms, as discussed in \cite{Banados:2022nhj, upcoming}. 

\section{From bulk diffeomorphism to conformal invariance} \label{sec: diffeo}

In this section we present an alternative derivation of \eqref{K_tranf} that makes clear how conformal symmetry emerges from bulk diffeomorphism. This derivation also easily extends to general fields and 
we discuss the case of gauge fields. 

Recall that the boundary field $\varphi_{(0)}(\vec{x})$
that parametrizes the boundary condition of a bulk scalar field $\phi (x)$ is given by
\begin{equation}
    \varphi_{(0)}(\vec{x}) = \lim_{z\rightarrow0}\ z^{\Delta-d}\phi(x)\,.
\end{equation}
As we discussed in section \ref{sec:AdSIso} the radial coordinate under the isometry transformations in \eqref{is1}-\eqref{is4} transforms as,
\begin{equation} \label{z_tr}
    z' = \tilde{\Omega}(x) z
\end{equation}
where $\tilde{\Omega}(x)$ has the property 
\begin{equation} \label{OmegaB_lim}
  \lim_{z\to 0} \tilde{\Omega}(x)= \Omega(\vec{x}),  
\end{equation}
with $\Omega(\vec{x})$ the Jacobian factor of the conformal transformations listed in \eqref{conf_1}-\eqref{conf_4}.
Then, adapting an argument from \cite{Bianchi:2001de, Kostas}, 
we find that the source transforms as follows
\begin{equation} \label{sc_sou_tr}
    \varphi_{(0)}'(\vec{x}\pr) = \lim_{z'\rightarrow0}\ z'^{\Delta-d}\phi'(x')=
   % \,,\\&= 
    \lim_{z\rightarrow0}\ \tilde{\Omega}^{\Delta-d}(x)z^{\Delta-d}\phi(x)= \Omega^{\Delta-d}(\vec{x})\varphi_{(0)}(\vec{x})\,,
\end{equation}
where in the second equality we used the fact that $\phi$ is a scalar under bulk diffeos and equation \eqref{z_tr}, and in the last equality we used \eqref{OmegaB_lim}. This is the expected transformation rule for a source that couples to a scalar operator of dimension $\Delta$. Now, at linearized order in the sources the bulk field is given by
\begin{equation} \label{Sc_lin}
    \phi(x)=\int\dd^dy\ K_\Delta(x,\vec{y})\varphi_{(0)}(\vec{y})
\end{equation}
Thus,
\begin{align} 
    \phi'(x') &= \int\dd^dy'\ K_\Delta(x',\vec{y}\pr)\varphi'_0(\vec{y}\pr)\,, \nonumber \\
    &= \int\dd^dy\ \Omega^d(\vec{y})K_\Delta(x',\vec{y}\pr)\Omega^{\Delta-d}(\vec{y})\varphi_{(0)}(\vec{y})\,, \nonumber \\
    &= \int\dd^dy\ \Omega^\Delta(\vec{y})K_\Delta(x',\vec{y}\pr)\varphi_{(0)}(\vec{y})\,, \label{phiprime}
\end{align}
Since this a scalar field, $\phi'(x')=\phi(x)$, and comparing \eqref{Sc_lin} with \eqref{phiprime} we conclude that the bulk-to-boundary propagator transforms as a scalar primary field:
\begin{equation}
    K_\Delta(x',\vec{y}\pr) = \Omega^{-\Delta}(\vec{y})K_\Delta(x,\vec{y})\,,
\end{equation}
Another way to see this is to note that \eqref{Sc_lin} has the same form as the coupling of the source to the operator $\int\dd^dy\ {\cal O}(\vec{y}) \phi_{(0)}(\vec{y})$. 

\subsection{Generalization to spinning operator}

This discussion readily generalises to spinning fields. The higher the spin the more complex the formulas and to keep the technicalities to the minimum we will present the details for a gauge field. All the steps, however, are the same in all cases.  As in the case of a scalar, the first step is to establish that the sources indeed transforms as a source of a spinning primary operator. For a gauge field the source is given by
\begin{equation}
    a_{(0)\alpha}(\vec{x}) = \lim_{z\rightarrow0}\ A_\alpha(x)\,.
\end{equation}
We now follow the same steps as in \eqref{sc_sou_tr}: 
\begin{align}
    a'_{(0)\alpha}(\vec{x}\pr) &= \lim_{z'\rightarrow0}\ A'_\alpha(x')
    = \lim_{z\rightarrow0}\ \frac{\partial x^\mu}{\partial x'^\alpha}A_\mu(x)
    = \lim_{z\rightarrow0}\ \tilde{\Omega}^{-1}(x)\tilde{R}_\alpha^{\ \mu}(x)A_\mu(x) \nonumber \\
    &= \Omega^{-1}(\vec{x})R_\alpha^{\ \beta}(\vec{x})a_{(0)\beta}(\vec{x})\,, \label{vec_trans}
\end{align}
where we used \eqref{adsjacob}, \eqref{Jac_limit} and the fact that the radial component of the field, $A_z$, is subleading in $z$ and thus vanishes as $z \to 0$. This is indeed the correct transformation for a source that couples to a conserved current of dimension $\Delta=d-1$.

The bulk gauge field to linear order in the sources is given by
\begin{equation}
    A_\mu(x) =\int\dd^dy\ K_\mu^{\ \alpha}(x,\vec{y})a_{(0) \alpha}(\vec{y})\, .
\end{equation}
Following the same steps as in \eqref{phiprime} a
we find:
\begin{align}
    A'_\mu(x') &= \int\dd^dy'\ K_\mu^{\ \alpha}(x',\vec{y}\pr)a'_{(0)\alpha}(\vec{y}\pr)
    = \int\dd^dy\ \Omega^d(\vec{y})K_\mu^{\ \alpha}(x',\vec{y}\pr)\Omega^{-1}(\vec{y})R_\alpha^{\ \beta}(\vec{y})a_{(0)\beta}(\vec{y})\nonumber \\
    &= \int\dd^dy\ \Omega^{d-1}(\vec{y})K_\mu^{\ \alpha}(x',\vec{y}\pr)R_\alpha^{\ \beta}(\vec{y})a_{(0)\beta}(\vec{y})\,.
\end{align}
By diffeomorphism invariance:
\begin{equation} \label{eq: K_trasf}
   A'_\mu(x') =  \frac{\partial x^\nu}{\partial x'^\mu}A_\nu (x)\quad \Rightarrow \quad
     K_{\mu\alpha}(x',\vec{y}\pr) = \Omega^{-(d-1)}(\vec{y})\frac{\partial x^\nu}{\partial x'^\mu}K_{\nu\beta}(x,\vec{y})R_\alpha^{\ \beta}(\vec{y})\,,
\end{equation}
and we reproduce \eqref{Kvec_tra}.

A general AdS amplitude of $n$ conserved currents is given by
\begin{equation}
    I_{\alpha_1\dotsc\alpha_n}(\vec{y}_1,\dotsc,\vec{y}_n) = \int_{x_1}\dotsm\int_{x_n}K_{\mu_1\alpha_1}(x_1,\vec{y}_1)\dotsm K_{\mu_n\alpha_n}(x_n,\vec{y}_n)G^{\mu_1\dotsm\mu_n}(x_1,\dotsc,x_n)\,,
\end{equation}
where $G^{\mu_1\dotsm\mu_n}(x_1,\dotsc,x_n)$ is the amputated bulk $n$-point function of the gauge field $A_\mu$ (to any loop order).
Provided it transforms under diffeomorphisms as indicated by its indices 
\begin{equation}
    G^{\mu_1\dotsm\mu_n}(x'_1,\dotsc,x'_n) = \frac{\partial x'^{\mu_1}_1}{\partial x^{\nu_1}_1}\dotsm\frac{\partial x'^{\mu_n}_n}{\partial x^{\nu_n}_n}G^{\nu_1\dotsm\nu_n}(x_1,\dotsc,x_n)\,,
\end{equation}
a straightforward computation shows that the amplitudes transform as
\begin{equation}
    I_{\alpha_1\dotsc\alpha_n}(\vec{y}\pr_1,\dotsc,\vec{y}\pr_n) = \Omega^{-(d-1)}(\vec{y}_1)\dotsm\Omega^{-(d-1)}(\vec{y}_n)R_{\alpha_1}^{\ \beta_1}(\vec{y}_1) \cdots R_{\alpha_n}^{\ \beta_n}(\vec{y}_n)I_{\beta_1\dotsc\beta_n}(\vec{y}_1,\dotsc,\vec{y}_n)\,,
\end{equation}
which is indeed the transformation property of a CFT $n$-point function of conserved currents, see \eqref{eq:npt_vec}.

\section{Conclusions} \label{sec: conclusion}

We have shown that AdS amplitudes satisfy the conformal Ward identities and we obtained explicit formulas that compute the constants and functions of cross-ratios that appear in the CFT correlators in terms of bulk quantities. These are given in 
\eqref{c2}, \eqref{c3}, \eqref{eq:c4}, \eqref{cn} for scalar $n$-point functions. The same analysis can be carried out for spinning operators and we worked out explicitly the case of conserved currents. The constraints of conformal invariance originate from diffeomorphsim invariance in the bulk. 

Altogether these results imply that the AdS gravity is a CFT, but they do not yet imply that it is a local CFT. Local CFTs have local UV divergences, and thus the corresponding  bulk IR divergences should also be local. This has been established at tree-level in \cite{Henningson:1998gx, deHaro:2000vlm, Papadimitriou:2004ap, Bzowski:2015pba} and for scalar fields in AdS up to two loops in \cite{Banados:2022nhj, upcoming}. In addition, the conformal anomalies should  be that of a local CFT, and they are \cite{Henningson:1998gx, deHaro:2000vlm}. One should contrast these results with the case of de Sitter, where the bulk isometries also match that of (Euclidean) CFT. The Ward identities due to de Sitter isometries also take the form of conformal Ward identities, but the IR divergences of de Sitter in-in correlators and corresponding anomalies only partially match that of a local CFT \cite{Bzowski:2023nef}. Local CFTs are further constrained by OPEs. These may be used to express 4- and higher-point functions in terms of CFT data: conformal dimensions (encoded in 2-point functions) and OPE coefficients (encoded in 3-point functions, and these should satisfy bootstrap equations. We note that the functions of cross-ratios that appear in our analysis have special dependence on some of the cross-ratios (see comments below \eqref{inner} and \eqref{inner2}) and it would be interesting to understand the implication of this in the context of the bootstrap program.

The connection between CFT correlators and AdS amplitudes depends on the amputated bulk correlators transforming properly under bulk diffeomorphism. Such tranformation properties could be invalidated by UV and/or IR divergences. We used an AdS invariant regulator to ensure that UV issues do not cause any problems, but full details have only been worked out for scalar fields. It would be interesting to work out the regularised bulk-to-bulk propagators for general spinning field. At loop order and for gauge field the analysis would likely require to properly take into account the contribution of ghost fields. IR divergences do break (part of) the AdS isometries, but this breaking is linked to conformal anomalies and anomalous dimensions and it is a feature, not a problem. 
We also note any AdS covariant $n$-point function, irrespectively of how it is obtained would automatically yield a solution of the CFT Ward identities -- here we assumed they are computed by bulk perturbation theory, but {\it a priori} there could be other non-perturbative constructions.

{\bf Acknowledgement:} 
KS is supported in part by the STFC consolidated grant ST/T000775/1 “New Frontiers in Particle Physics, Cosmology and Gravity”. 

%\printbibliography
\bibliographystyle{ieeetr}
\bibliography{main}

\end{document}